\pdfoutput=1
\documentclass[journal,twocolumn,10pt,twoside]{IEEEtran}

\normalsize
\usepackage[pdftex]{graphicx}
\DeclareGraphicsExtensions{.pdf,.jpeg,.png}

\usepackage{amsmath}

\usepackage{array}

\usepackage{subfigure} 

\usepackage{amssymb}

\usepackage{dsfont}

\usepackage{verbatim}

\usepackage{amsthm}

\usepackage{stfloats}

\newtheorem{my_prop}{Proposition}

\global\long\def\rb#1{\left(#1\right)}
\global\long\def\sb#1{\left[#1\right]}
\global\long\def\cb#1{\left\{  #1\right\}  }

\global\long\def\abs#1{\left|#1\right|}

\newcommand{\given}[2]{\left.#1\right|#2}

\newcommand{\Text}[1]{\text{\textnormal{#1}}}

\begin{document}

\setlength{\IEEEilabelindent}{\IEEEilabelindentB}

\newcounter{MYtempeqncnt}

\title{Robust Uplink Communications over Fading Channels with Variable Backhaul Connectivity}

\author{Roy~Karasik,~\IEEEmembership{Student~Member,~IEEE,}
        Osvaldo~Simeone,~\IEEEmembership{Member,~IEEE,}
        and~Shlomo~Shamai~(Shitz),~\IEEEmembership{Fellow,~IEEE}
\thanks{R. Karasik and S. Shamai are with the Faculty of Electrical Engineering, Technion-Israel
Institute of Technology, Haifa 32000, Israel (e-mail: royk@tx.technion.ac.il).}
\thanks{O. Simeone is with the Center for Wireless Communications and Signal
Processing Research, New Jersey Institute of Technology, Newark, NJ, 07102-1982 USA (e-mail: osvaldo.simeone@njit.edu).}}

\markboth{IEEE Transactions on Wireless Communications}%
{IEEE Transactions on Wireless Communications}

\maketitle

\begin{abstract}
Two mobile users communicate with a central decoder via two base stations. Communication between the mobile users and the base stations takes place over a Gaussian interference channel with constant channel gains or quasi-static fading. Instead, the base stations are connected to the central decoder through orthogonal finite-capacity links, whose connectivity is subject to random fluctuations. There is only receive-side channel state information, and hence the mobile users are unaware of the channel state and of the backhaul connectivity state, while the base stations know the fading coefficients but are uncertain about the backhaul links' state. The base stations are oblivious to the mobile users' codebooks and employ compress-and-forward to relay information to the central decoder. Upper and lower bounds are derived on average achievable throughput with respect to the prior distribution of the fading coefficients and of the backhaul links' states. The lower bounds are obtained by proposing strategies that combine the broadcast coding approach and layered distributed compression techniques. The upper bound is obtained by assuming that all the nodes know the channel state.
Numerical results confirm the advantages of the proposed approach with respect to conventional non-robust strategies in both scenarios with and without fading.
\end{abstract}


\IEEEpeerreviewmaketitle

\section{Introduction}
\begin{figure}[!t]
\centering
\includegraphics[scale=0.5]{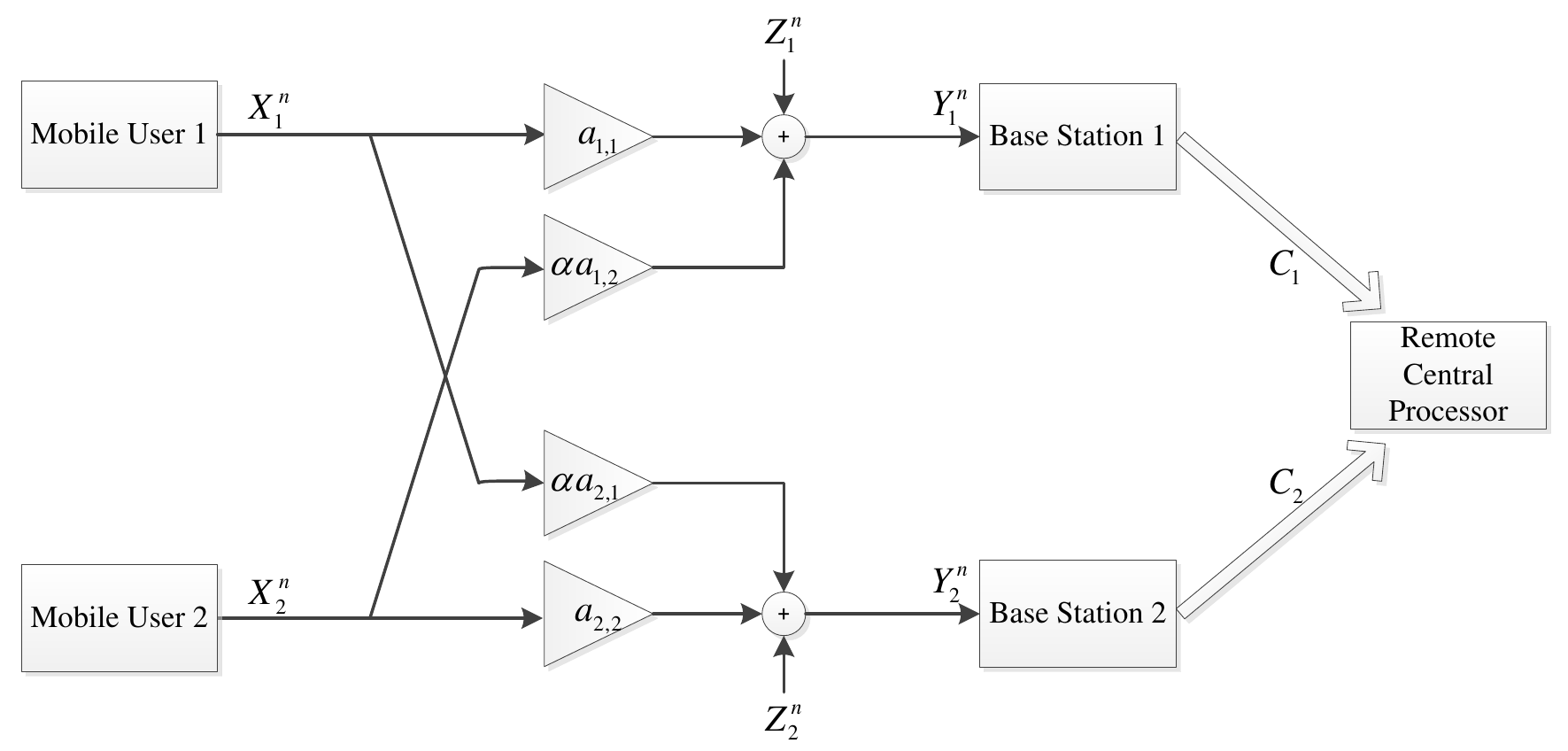}
\caption{Two-cell Gaussian cellular uplink channel with variable capacity backhaul links.}
\label{fig:system}
\end{figure}

\IEEEPARstart{M}{odern} cellular communication systems that implement the idea of network MIMO \cite{Gesbert} can be modeled by two-hop channels. Considering the uplink and with reference to Fig. \ref{fig:system}, the first hop corresponds to the channels between the Mobile Users (MUs) and the Base Stations (BSs), while the second hop accounts for communication between the BSs and the Remote Central Processor (RCP) that performs decoding across all connected cells. The first hop is generally to be regarded as a fading interference channel, capturing the wireless connection between MUs and BSs. Instead, the second hop can be often modeled by orthogonal wireless or wired backhaul links between each BS and the RCP. We refer to \cite{Gesbert},\cite{Foundation} for extensive reviews of the literature on network MIMO.

A specific implementation of network MIMO that is becoming of increasing interest is the so called cloud radio access network, whereby the BSs act as ``soft'' relays towards the destination (see, .e.g, \cite{Marsch}). In information-theoretic terms, the relays perform compress-and-forward in the second hop. Note that this system has the advantage that the relays do not need to be informed about the codebooks used by the MUs. The performance of the uplink system of Fig. \ref{fig:system}, and close variants, with compress-and-forward relays has been studied in \cite{Coso} and \cite{Park} assuming non-fading channels, and in \cite{Sanderovich2} assuming non-fading or ergodic fading channels in the first hop and backhaul links of given capacity in the second hop. Ergodic fading channels in the first hop are also considered in \cite{SanderovichMIMO} for a single-MU system with several antennas. The case of a single MU system was also considered in \cite{Simeone} but under the different assumption that the first hop has no fading but the backhaul links in the second hop may be in outage, unbeknownst to the MUs and to the BSs. Finally, reference \cite{Zamani} studies the single-MU case under the assumption that both the first and the second hop are subject to quasi-static fading.

In this paper, we revisit the system model in Fig. \ref{fig:system} by assuming both non-fading and quasi-static fading channels for the first hop and backhaul links with variable connectivity. Specifically, the backhaul links are assumed to be in one of two possible states, hence modeling in a simple fashion either wireless or wired links with variable quality. The main focus of this paper is on applications in which MUs and BSs have to operate with only receive but not transmit channel state information. In particular, the MUs are assumed to be aware neither of the fading channels nor of the backhaul links' state, while the BSs are not informed about the state of the backhaul links. This is the case for instance in low-delay applications in which it is not possible to accommodate feedback to the MUs and BSs.

As in \cite{Simeone},\cite{Zamani}, the MUs cope with the lack of channel state information by accepting variable-rate data delivery via broadcast coding (BC) \cite{Shamai}. Moreover, in order to opportunistically leverage better backhaul conditions, we resort to layered compression at the relays as in \cite{Simeone}. Overall, the proposed approach can be seen as an extension of \cite{Simeone} to a set-up with multiple users, fading channels in the first hop and a more general backhaul state model. Upper and lower bounds are derived on average achievable throughput with respect to the prior distribution of the fading coefficients and of the backhaul links' states. As mentioned, the lower
bounds are obtained by proposing strategies that combine the BC approach and layered distributed compression techniques \cite{Rimoldi},\cite{Chen}. The upper bound is instead obtained by relaxing the system's constraints, namely by assuming that all the nodes know the channel state, and leveraging the results of \cite{Sanderovich1}.

The rest of the paper is organized as follows. We introduce the model in Section \ref{sec:2}. Section \ref{sec:3} contains lower and upper bounds for the average throughput without fading in the first hop, i.e., only variable backhaul connectivity is considered. In Section \ref{sec:4} the achievable region is generalized to the case of Rayleigh quasi-static fading. Numerical examples are presented in Section \ref{sec:5}. Finally, we make some concluding remarks in Section \ref{sec:6}.

\textit{Notation:} $\mathbf{I}$ denotes the $2\times 2$ identity matrix, and $\mathds{1}_{\cb{j=1}}$ equals $1$ for $j=1$ and $0$ otherwise. $\text{diag}\rb{a,b}$ represents a diagonal matrix with main diagonal $(a,b)$. Gaussian and circularly-symmetric complex Gaussian random variables, with mean $\mu$ and variance $\sigma^2$, are denoted by $\mathcal N(\mu,\sigma^2)$ and $\mathcal C\mathcal N(\mu,\sigma^2)$, respectively. Moreover, using standard notation, we will sometimes use superscripts to denote index bounds in sequences as in $x_1^n=\rb{x_{1,1},\ldots,x_{1,n}}$. The use of the superscript will be made clear by the context. Finally, for $j\in\mathbb Z$, $[j]_2\triangleq j\text{ mod }2$.

\section{System Model and Preliminaries}\label{sec:2}
We consider the uplink cellular multiple access model of Fig. \ref{fig:system} that consists of two identical cells, indexed by $j=1,2$. Each cell includes a single-antenna MU and a single-antenna BS. The MUs wish to send information to a RCP, using the BSs as relay stations.
Each $j$th BS receives the transmission $X_j$ of its cell's MU with superimposed interference from the MU of the adjacent cell and independent white Gaussian noise $Z_j$.
The received signals at the BSs for time index $i$ read
\begin{IEEEeqnarray}{rCl}\label{eq:sys_mod}
  Y_{1,i}&=&a_{1,1}X_{1,i}+\alpha a_{1,2}X_{2,i}+Z_{1,i},\IEEEyessubnumber\\
  Y_{2,i}&=&\alpha a_{2,1}X_{1,i}+a_{2,2}X_{2,i}+Z_{2,i}\IEEEyessubnumber
\end{IEEEeqnarray}
for $i=1,2,\ldots,n$, where $\rb{a_{1,1},a_{1,2},a_{2,1},a_{2,2}}$ represent the channel gains, and the inter-cell interference path loss coefficient is $0\leq\alpha\leq 1$. Each $j$th MU has an average power constraint $\frac{1}{n}\sum_{i=1}^{n}|x_{j,i}|^2\leq P$ for $j=1,2$.

Two scenarios will be considered in the paper. The first consists of \textit{non-fading} channels, that is, we have $a_{1,1}=a_{1,2}=a_{2,1}=a_{2,2}=1$. In this case, it is assumed that the symbols $X_{1,i}$ and $X_{2,i}$ are real and that the noise is distributed as $Z_{j,i}\sim \mathcal N(0,1)$. This assumption is made without loss of generality since the in-phase and quadrature components of the equivalent baseband signal can be treated separately in the absence of fading. In the second scenario, Rayleigh \textit{quasi-static fading} channels are assumed, and thus the channel coefficients $a_{1,1},a_{1,2},a_{2,1}$ and $a_{2,2}$ are independent, distributed as $\mathcal C\mathcal N(0,1)$, and constant during the transmission block.
The MUs are unaware of the realization of the fading coefficients, as opposed to the BSs and the RCP that have instead full channel state information. The transmitted signals $X_{1,i}$ and $X_{2,i}$ are complex and the noise is distributed $Z_{j,i}\sim \mathcal C\mathcal N(0,1)$.

The BSs are assumed to be unaware of the codebooks used by the MUs, as was assumed in \cite{Sanderovich2}, \cite{Simeone} and \cite{Sanderovich1},
and are connected to the RCP via orthogonal finite-capacity backhaul links, e.g., dedicated wireless or wired connections.
The connectivity of the backhaul links is uncertain in the sense that each link can have two possible states: a \textit{low-capacity} state, in which the capacity is $C$, and a \textit{high-capacity} state, in which the capacity is $C+\Delta C$ with $\Delta C\geq 0$. As a result, for $j=1,2$, the capacity $C_j$ of the $j$th backhaul link is given by random variable
\begin{IEEEeqnarray}{rCl}
    C_j=\left\lbrace\begin{array}{ll}C,& \mbox{with probability }p\\ C+\Delta C,& \mbox{with probability }1-p\end{array}\right.\label{eq:sys_C}
\end{IEEEeqnarray}
for some $0\leq p\leq 1$. Moreover, $C_1$ is independent of $C_2$. Also, the state of the backhaul links remains constant in the communication block, and no instantaneous information regarding the current state of the backhaul links, i.e., the value of $C_j$, is available to the MUs and the BSs. The RCP, instead, is aware of the current state of the backhaul links.

We observe that, overall, our assumptions on the information about the state of the fading channels and backhaul links can be summarized by saying that full information is assumed at the receiver side (and at downstream nodes) and no information at the transmitter side (and at upstream nodes).

The $j$th BS compresses the received signal $Y_j^n$ to produce the indices $s_{j}\in\cb{1,2,\ldots,2^{nC}}$ and $r_{j}\in\cb{1,2,\ldots,2^{n\Delta C}}$ that are transmitted through the $j$th backhaul link. When the $j$th backhaul link is in the low-capacity state ($C$), the RCP receives only the $s_{j}$ index, and, when it is in the high-capacity state ($C+\Delta C$), both indices $s_{j}$ and $r_j$ are received by the RCP.

In order to combat fading and the uncertainty of the backhaul links, the MUs employ the BC approach \cite{Shamai}.
Accordingly, each MU divides its information message $M_j$ into $K$ independent  sub-messages $M_{j}=\rb{M_{j,1},M_{j,2},\ldots,M_{j,K}}$, $j=1,2$. Let $R_{j,k}$ be the rate of the $k$th message, $k=1,2,\ldots,K$, i.e., $M_{j,k}\in\cb{1,2,\ldots,2^{nR_{j,k}}}$. Moreover, for a given channel-backhaul realization $\rb{\mathbf{a},\mathbf{c}}\triangleq\rb{a_{1,1},a_{1,2},a_{2,1},a_{2,2},c_1,c_2}$, where $c_j\in\cb{C,C+\Delta C}$, let $\mathcal I_{\rb{\mathbf{a},\mathbf{c}}}$ be the set of indices of messages that can be decoded by the RCP, that is
\begin{IEEEeqnarray}{rl}
  &\mathcal I_{\rb{\mathbf{a},\mathbf{c}}}=
  \cb{\given{\rb{j,k}\in\mathcal A_{J,K}}{M_{j,k}\text{ is decodable given }\rb{\mathbf{a},\mathbf{c}}}}\IEEEeqnarraynumspace
\end{IEEEeqnarray}
where $\mathcal A_{J,K}\triangleq \cb{1,2}\times\cb{1,2,\ldots,K}$.
This set depends on the specific coding/decoding strategy, as it will be discussed. Finally, define the throughput given the channel-backhaul realization $\rb{\mathbf{a},\mathbf{c}}$ as
\begin{IEEEeqnarray}{rCl}
  T_{\rb{\mathbf{a},\mathbf{c}}}=\sum_{(j,k)\in \mathcal I_{\rb{\mathbf{a},\mathbf{c}}}}R_{j,k}.\label{eq:throughput}
\end{IEEEeqnarray}
The performance criterion of interest is the average achievable throughput $T$, where the average is taken with respect to the a priori probability of fading coefficients and backhaul link state.
We also remark that, as in \cite{Simeone}, the average throughput $T$ does not have the operational meaning of ergodic sum-rate, since the channel is nonergodic. Instead, the average throughput stands for the average sum-rate that can be accrued with repeated and independent transmission block, in the long run, or for the expected throughput.

\section{No Fading}\label{sec:3}
\begin{figure*}[!t]
  \normalsize
  \setcounter{MYtempeqncnt}{\value{equation}}
  \setcounter{equation}{5}
  {\normalsize{
  \begin{IEEEeqnarray}{rCl}\label{eq:nf_cond}
    R_{1,1}+R_{2,1}&<& \frac{1}{2}\log\det\rb{\mathbf{I}+\lambda_1 P\rb{\mathbf{A}_1+\mathbf{A}_2}
    \sb{\mathbf{\Lambda}_I\rb{\cb{2,3,4,5},\cb{2,3,4,5}}+\rb{1+\sigma_1^2+\sigma_2^2}\mathbf{I}}^{-1}}\IEEEyessubnumber\\
    R_{1,2}+R_{2,2}&<&\frac{1}{2}\log\det\rb{\mathbf{I}+\lambda_2 P\rb{\mathbf{A}_1+\mathbf{A}_2}
    \sb{\mathbf{\Lambda}_I\rb{\cb{3,4,5},\cb{3,4,5}}+\Text{diag}\rb{1+\sigma_2^2,1+\sigma_1^2+\sigma_2^2}}^{-1}},\IEEEyessubnumber\\
    R_{j,3}&<& \frac{1}{2}\log\det\rb{\mathbf{I}+\lambda_3 P\mathbf{A}_1
    \sb{\mathbf{\Lambda}_I\rb{\cb{4,5},\cb{3,5}}+\Text{diag}\rb{1+\sigma_2^2,1+\sigma_1^2+\sigma_2^2}}^{-1}},\IEEEyessubnumber\\
    R_{j,4}&<& \frac{1}{2}\log\det\rb{\mathbf{I}+\lambda_4 P\mathbf{A}_2
    \sb{\mathbf{\Lambda}_I\rb{\cb{4,5},\cb{3,5}}+\Text{diag}\rb{1+\sigma_2^2,1+\sigma_1^2+\sigma_2^2}}^{-1}},\IEEEyessubnumber\\
    R_{j,3}+R_{[j]_2+1,4}&<& \frac{1}{2}\log\det\rb{\mathbf{I}+P\rb{\lambda_3\mathbf{A}_1+\lambda_4\mathbf{A}_2}
    \sb{\mathbf{\Lambda}_I\rb{\cb{4,5},\cb{3,5}}+\Text{diag}\rb{1+\sigma_2^2,1+\sigma_1^2+\sigma_2^2}}^{-1}},\IEEEyessubnumber\\
    R_{1,5}+R_{2,5}&<& \frac{1}{2}\log\det\rb{\mathbf{I}+\lambda_5 P\rb{1+\sigma_2^2}^{-1}\rb{\mathbf{A}_1+\mathbf{A}_2}
    },\IEEEyessubnumber
  \end{IEEEeqnarray}
  }}
  \setcounter{equation}{\value{MYtempeqncnt}}
  \hrulefill
  \vspace*{4pt}
\end{figure*}
In this section, the special case where there is no fading is considered.
Without fading, the only uncertainty of the MUs and BSs is on the state of the backhaul links. In fact, there are four possible states of the backhaul links $\rb{C_1,C_2}$, namely $\rb{C,C},\rb{C+\Delta C,C},\rb{C,C+\Delta C}$ and $\rb{C+\Delta C,C+\Delta C}$, which will be labeled as state 1, state 2, state 3 and state 4, respectively.
We denote by ``Decoder $l$'' the decoding scheme used by the RCP in state $l$, $l=1,2,3,4$. Note that Decoder 1 receives the indices $\rb{s_1,s_2}$ from the BSs, while Decoder 2 receives $\rb{s_1,s_2,r_1}$, Decoder 3 receives $\rb{s_1,s_2,r_2}$, and Decoder 4 receives $\rb{s_1,s_2,r_1,r_2}$. As a result, Decoder 2 and Decoder 3 can decode all the MUs' messages that Decoder 1 can decode, and Decoder 4 can decode any messages that the other decoders are capable of decoding. However, Decoder 2 might not be able to decode messages that Decoder 3 can decode, and vice versa, since they receive different subsets of indices from the BSs. This situation contrasts with the (non-fading) single-user model studied in \cite{Simeone}, in which, due to the symmetry of the system model, the states could be ordered depending on their decoding power.
We now propose a strategy to cope with the problem outlined above and then present an upper bound on the achievable throughput.
\subsection{Achievable Throughput}\label{subsec:NF_achievable}
\begin{figure}[!t]
\centering
\includegraphics[scale=0.6]{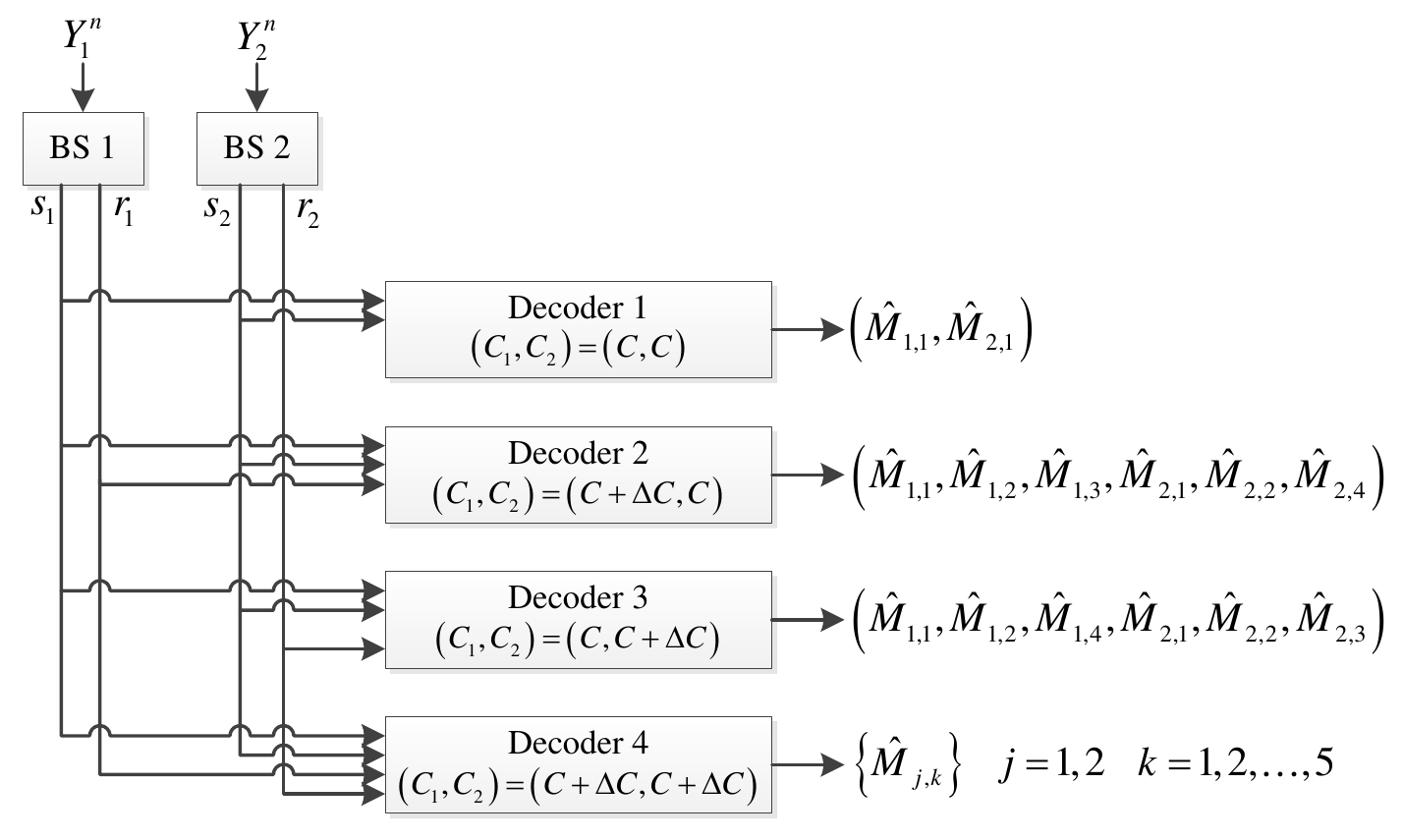}
\caption{Illustration of the decoding strategy for the proposed BC scheme for the non-fading model.}
\label{fig:Drawing}
\end{figure}

To deal with the backhaul uncertainty, we adopt the broadcast strategy proposed in \cite{Whiting} in the context of broadcasting under delay constraints. Accordingly, both MUs use five messages ($K=5$): messages $\rb{M_{1,1},M_{2,1}}$ are to be decoded by the RCP no matter what the backhaul state is, and hence by Decoder 1, 2, 3, 4; $\rb{M_{1,2},M_{2,2}}$ are to be decoded when either $C_1=C+\Delta C$ or $C_2=C+\Delta C$, and hence by Decoder 2, 3 and 4; $\rb{M_{1,3},M_{2,4}}$ are to be decoded whenever $C_1=C+\Delta C$, and hence by Decoder 2 and 4; $\rb{M_{1,4},M_{2,3}}$ are to be decoded whenever $C_2=C+\Delta C$, and hence by Decoder 3 and 4; and $\rb{M_{1,5},M_{2,5}}$ are to be decoded when both $C_1=C+\Delta C$ and $C_2=C+\Delta C$, and hence only by Decoder 4.
The assignment of messages and decoders is illustrated in Fig. \ref{fig:Drawing}.
As a result of these choices, \eqref{eq:throughput} can be written as: $T_1\triangleq T_{\rb{C,C}}=\sum_{j=1}^{2}R_{j,1}$, $T_2\triangleq T_{\rb{C+\Delta C,C}}=\sum_{j=1}^{2}\sum_{k=1}^{2}R_{j,k}+R_{1,3}+R_{2,4}$, $T_3\triangleq T_{\rb{C,C+\Delta C}}=\sum_{j=1}^{2}\sum_{k=1}^{2}R_{j,k}+R_{1,4}+R_{2,3}$ and $T_4\triangleq T_{\rb{C+\Delta C,C+\Delta C}}=\sum_{j=1}^{2}\sum_{k=1}^{5}R_{j,k}$, and 
the average throughput $T$ is thus
\begin{IEEEeqnarray}{rCl}
    T&=& p^2T_1+p(1-p)\left(T_2+T_3\right)+(1-p)^2T_4.\label{eq:nf_average}
\end{IEEEeqnarray}

Decoding at the RCP is performed as follows. As mentioned, the $j$th BS sends two indices, $s_j$ and $r_j$, on the backhaul link. These are obtained by compressing the received signal $Y_j^n$ using a layered quantization codebook: the base layer provides a coarse description of $Y_j^n$ and is encoded only in the index $s_j$, while the overall codebook provides a refined description of $Y_j^n$ and is encoded by both indices $\rb{s_j,r_j}$. The rate of the coarse description is $C$ bits, while the rate of the refined description is $\rb{C+\Delta C}$ bits.
The RCP first recovers the compressed versions of the received signals $Y_j^n$, either coarse or refined, depending on the backhaul links' state. These compressed received signals are used by the RCP to decode the messages corresponding to the current state of the backhaul links, as illustrated in Fig. \ref{fig:Drawing}.

Overall, the proposed strategy is based on BC coding at the MUs and layered quantization at the BSs. The scheme generalizes the approach proposed in \cite[Sec.~IV]{Simeone}, where the focus was on (non-fading) single-user systems with $C=0$. The proposed approach achieves the average throughput described in the proposition below.
\begin{my_prop}[BC and separate decompression]\label{prop:achievable}
  Let $\lambda_1,\lambda_2,\lambda_3,\lambda_4,\lambda_5\geq0$ such that $\sum_{l=1}^{5}\lambda_l=1$.
  The average throughput \eqref{eq:nf_average} is achievable with \addtocounter{equation}{1} \eqref{eq:nf_cond} at the top of the next page, where
  \begin{IEEEeqnarray}{rCl}
    \mathbf{A}_1\triangleq\rb{\begin{array}{cc}1&\alpha\\ \alpha&\alpha^2\end{array}}\text{ and }
    \mathbf{A}_2\triangleq\rb{\begin{array}{cc}\alpha^2&\alpha\\ \alpha&1\end{array}},
  \end{IEEEeqnarray}
  \begin{IEEEeqnarray}{rCl}\label{eq:nf_sig}
        \sigma_1^2&=&\frac{2^{2C}\rb{2^{2\Delta C}-1}\rb{P(1+\alpha^2)+1}}{\rb{2^{2C}-1}\rb{2^{2(C+\Delta C)}-1}},\IEEEyessubnumber\\
        \sigma_2^2&=&\frac{P(1+\alpha^2)+1}{2^{2(C+\Delta C)}-1},\IEEEyessubnumber
  \end{IEEEeqnarray}
   and for $\mathcal I_1,\mathcal I_2\subseteq\{1,2,3,4,5\}$
  \begin{IEEEeqnarray}{rl}\label{eq:interference_layers}
    \mathbf{\Lambda}_I&\rb{\mathcal I_1,\mathcal I_2}\triangleq
    P\sb{\rb{\sum_{k\in\mathcal I_1}\lambda_k}\mathbf{A}_1+\rb{\sum_{k\in\mathcal I_2}\lambda_k}\mathbf{A}_2}.
  \end{IEEEeqnarray}
\end{my_prop}
Proposition \ref{prop:achievable} is proved in Appendix \ref{app:NF_achievable} by analyzing the performance of the strategy summarized above. A brief discussion is presented here, along with an interpretation of the parameters that appear in Proposition \ref{prop:achievable}. The MUs use BC based on i.i.d. generated Gaussian codebooks, where parameters $\rb{\lambda_1,\lambda_2,\ldots,\lambda_5}$ represent the power allocation used by both MUs to transmit the five messages $M_{j,1},\ldots,M_{j,5}$. Specifically, each $j$th MU transmits a signal
\begin{IEEEeqnarray}{rCl}\label{eq:NF_BC}
    X_{j}=\sqrt{P}\sum_{k=1}^{5}\sqrt{\lambda_k}W_{j,k},
\end{IEEEeqnarray}
where $W_{j,k}\sim \mathcal N(0,1)$ for $k=1,\ldots,5$, are the independent variables representing the five codebooks used to encode the corresponding messages $M_{j,k}$ for $k=1,\ldots,5$. In every backhaul state, the layers $W_{j,k}$ that cannot be decoded by the RCP act as additional additive noise, as in classical BC \cite{Shamai}, while all the messages that are to be retrieved according to Fig. \ref{fig:Drawing} are jointly decoded. As a result, inequalities \eqref{eq:nf_cond} follow from capacity results on multiple access channels (see, e.g., \cite{KimGamal}), and matrix $\mathbf{\Lambda}_I\rb{\mathcal I_1,\mathcal I_2}$ in \eqref{eq:interference_layers} is the covariance matrix of the interfering signal corresponding to the uncoded messages when the messages $\cb{M_{j,k}}$ with $k\in\mathcal I_j$ and $j=1,2$ cannot be decoded in the current backhaul links state.

The quantities $\rb{\sigma_1^2,\sigma_2^2}$ in \eqref{eq:nf_sig} represent the quantization noise introduced by compression at the BSs. Specifically, the compression noise on the refined quantization codeword has variance $\sigma_2^2$, while the basic quantization codeword is characterized by compression noise variance $\sigma_1^2+\sigma_2^2$.
More specifically, the relationship between the received signals and the corresponding compressed versions are defined by so-called test channels (see, e.g., \cite{KimGamal}), which are assumed here to be characterized by Gaussian noises with the mentioned variances. Accordingly, the compressed signals for BS $j=1,2$ can be written as
\begin{IEEEeqnarray}{rCl}\label{eq:NF_test1}
  V_{j,1}&=&Y_{j}+Q_{j,1}+Q_{j,2}
\end{IEEEeqnarray}
for the coarse description and
\begin{IEEEeqnarray}{rCl}\label{eq:NF_test2}
  V_{j,2}&=&Y_{j}+Q_{j,2}
\end{IEEEeqnarray}
for the refined description, where $\cb{Q_{j,l}}_{l=1}^2$ represent the compression noises, which distributed as $Q_{j,1}\sim \mathcal N(0,\sigma_1^2)$ and $Q_{j,2}\sim \mathcal N(0,\sigma_2^2)$.

\begin{figure*}[!b]
  \normalsize
  \setcounter{MYtempeqncnt}{\value{equation}}
  \setcounter{equation}{15}
  \vspace*{4pt}
  \hrulefill
  {\normalsize{
  \begin{IEEEeqnarray}{rCl}\label{eq:NF_upper_sum}
    T_1^{(\text{ub})}&=&\frac{1}{2}\log\det\rb{\mathbf{I}+P\rb{1+\sigma_1^2}^{-1}\rb{\mathbf{A}_1+\mathbf{A}_2}
    }\IEEEyessubnumber\\
    T_2^{(\text{ub})}&=&\max_{\sigma_2^2,\sigma_3^2}\frac{1}{2}\log\det\rb{\mathbf{I}+P\rb{\mathbf{A}_1+\mathbf{A}_2}
    \sb{\Text{diag}\rb{1+\sigma_2^2,1+\sigma_3^2}}^{-1}}\IEEEyessubnumber\label{eq:T2}\\
    T_3^{(\text{ub})}&=&\frac{1}{2}\log\det\rb{\mathbf{I}+P\rb{1+\sigma_4^2}^{-1}\rb{\mathbf{A}_1+\mathbf{A}_2}
    }\IEEEyessubnumber
  \end{IEEEeqnarray}
  }}
  \setcounter{equation}{\value{MYtempeqncnt}}
\end{figure*}
\begin{figure*}[!b]
  \normalsize
  \setcounter{MYtempeqncnt}{\value{equation}}
  \setcounter{equation}{18}
  \vspace*{4pt}
  \hrulefill
  {\normalsize{
  \begin{IEEEeqnarray}{rCl}\label{eq:NF_upper_var2}
    C+\Delta C&\geq& \frac{1}{2}\log\rb{\frac{\rb{P(1+\alpha^2)+1+\sigma_2^2}\rb{P(1+\alpha^2)+1+\sigma_3^2}-4\alpha^2P^2}{\rb{P(1+\alpha^2)+1+\sigma_3^2}\sigma_2^2}}\IEEEyessubnumber\\
    C&\geq& \frac{1}{2}\log\rb{\frac{\rb{P(1+\alpha^2)+1+\sigma_2^2}\rb{P(1+\alpha^2)+1+\sigma_3^2}-4\alpha^2P^2}{\rb{P(1+\alpha^2)+1+\sigma_2^2}\sigma_3^2}}\IEEEyessubnumber\\
    2C+\Delta C&\geq& \frac{1}{2}\log\rb{\frac{\rb{P(1+\alpha^2)+1+\sigma_2^2}\rb{P(1+\alpha^2)+1+\sigma_3^2}-4\alpha^2P^2}{\sigma_2^2\sigma_3^2}}.\IEEEyessubnumber
  \end{IEEEeqnarray}
  }}
  \setcounter{equation}{\value{MYtempeqncnt}}
\end{figure*}
As detailed in Appendix \ref{app:NF_achievable}, the variances $\sigma_1^2$ and $\sigma_2^2$ in \eqref{eq:nf_sig} are derived by assuming separate decompression of the indices of the two BSs by the RCP. With separate decompression, the RCP recovers the two compressed received signals from the BSs in parallel without any joint decoding across the BSs. A more efficient compression/decompression strategy can leverage the fact that the signals received by the BSs are correlated. Specifically, we can allow the RCP to jointly decompress the basic descriptions encoded in indices $s_1$ and $s_2$ by using distributed source coding, or binning, on the basic descriptions at the BSs. Using distributed source coding and joint decompression allows the additive equivalent noise on the basic descriptions to be reduced with respect to Proposition \ref{prop:achievable} (see, e.g., \cite{Sanderovich1}). Note that in the proposed approach, distributed source coding is used only on the indices $s_1$ and $s_2$, which are received at the RCP for all backhaul states. Using distributed source coding across other indices would cause errors in the decompression process for all states in which the involved indices are not received (see \cite{Park} for related discussion). As proved in Appendix \ref{app:NF_achievable}, distributed source coding of indices $s_1$ and $s_2$ leads to following achievable throughput.
\begin{my_prop}[BC and joint decompression]\label{cor:achievable}
  Let $\lambda_1,\lambda_2,\lambda_3,\lambda_4,\lambda_5\geq0$ such that $\sum_{l=1}^{5}\lambda_l=1$.
  The average throughput \eqref{eq:nf_average} is achievable with \eqref{eq:nf_cond}, where $\sigma_1^2$ and $\sigma_2^2$ are positive solutions of the two equations
  \begin{IEEEeqnarray}{rCl}
    2C&=&\frac{1}{2}\log\rb{1+\frac{P(1-\alpha)^2+1}{\sigma_1^2+\sigma_2^2}}\IEEEnonumber\\
    && +\:\frac{1}{2}\log\rb{1+\frac{P(1+\alpha)^2+1}{\sigma_1^2+\sigma_2^2}}\label{eq:nf_sig2}
  \end{IEEEeqnarray}
  and
  \begin{IEEEeqnarray}{rCl}
    \Delta C&=&\frac{1}{2}\log\rb{1+\frac{\sigma_1^2}{\sigma_2^2}}\label{eq:nf_sig3}\IEEEnonumber\\
    && +\:\frac{1}{2}\log\rb{1-\frac{\sigma_1^2}{P(1+\alpha^2)+1+\sigma_1^2+\sigma_2^2}}.
  \end{IEEEeqnarray}
\end{my_prop}

\subsection{Upper Bound}\label{sec:NF_upper}
An upper bound on the average throughput is derived here by assuming that all the nodes, MUs, BSs and RCP, know the backhaul states. This bound will be used to asses the efficacy of the strategy proposed above to combat the uncertainty regarding the backhaul states at the MUs and BSs. The bound holds only within the class of strategies based on i.i.d. Gaussian codebooks for the MUs and compression/decompression using distributed source coding and Gaussian test channels at the BSs. This class includes the achievable strategies proposed above and is considered for its practical relevance and simplicity of analysis. Note that Gaussian signaling is not necessarily optimal, as shown in \cite{Sanderovich1} for a system with one MU and deterministic backhaul links, and that the optimality of Gaussian test channel was shown in related problems (e.g. \cite{Coso}), but need not hold here.
\begin{my_prop}\label{prop:upper}
  The following is an upper bound on the average throughput \eqref{eq:nf_average} for strategies based on i.i.d. Gaussian codebooks for the MUs and compression/decompression using distributed source coding and Gaussian test channels at the BSs.
  \begin{IEEEeqnarray}{rCl}\label{NF_upper_avg}
    T &\leq& p^2T^{(\text{ub})}_1+2p(1-p)T_2^{(\text{ub})}
     +\:(1-p)^2T_3^{(\text{ub})}
  \end{IEEEeqnarray}
  where $T_1^{(\text{ub})},T_2^{(\text{ub})}$ and $T_3^{(\text{ub})}$ are given by \addtocounter{equation}{1} \eqref{eq:NF_upper_sum}, at the bottom of the page,
  with $\sigma_1^2$ and $\sigma_4^2$ being the positive solutions of
  \begin{IEEEeqnarray}{rCl}
    2C&=&\frac{1}{2}\log\rb{1+\frac{P(1-\alpha)^2+1}{\sigma_1^2}}\IEEEnonumber\\
    &&+\:\frac{1}{2}\log\rb{1+\frac{P(1+\alpha)^2+1}{\sigma_1^2}}\label{eq:NF_cond1}
  \end{IEEEeqnarray}
  and
  \begin{IEEEeqnarray}{rCl}
    2C+2\Delta C&=&\frac{1}{2}\log\rb{1+\frac{P(1-\alpha)^2+1}{\sigma_4^2}}\IEEEnonumber\\
    &&+\:\frac{1}{2}\log\rb{1+\frac{P(1+\alpha)^2+1}{\sigma_4^2}}\IEEEeqnarraynumspace\label{eq:NF_cond2}
  \end{IEEEeqnarray}
  respectively, while the maximization in \eqref{eq:T2} is performed with respect to $\sigma_2^2,\sigma_3^2\geq0$ that satisfy \addtocounter{equation}{1} \eqref{eq:NF_upper_var2} at the bottom of the page.
\end{my_prop}
\begin{IEEEproof}
  Since every node is aware of the backhaul state, the throughput can be calculated separately for each state and then an average is taken with respect to the probability distribution of the states to obtain the average throughput \eqref{NF_upper_avg}. By assumption, the MUs transmit using independent Gaussian i.i.d. codebooks designed for the current backhaul state. In each state, the sum-rate must be less than the upper bound derived for the single-user system of \cite{Sanderovich1}. Applying \cite[Theorem 2]{Sanderovich1} with Gaussian inputs and test channels, we get Proposition \ref{prop:upper}. Specifically, $T_1^{(\text{ub})}$ and $T_3^{(\text{ub})}$ bound the sum-rate in states 1 and 4, respectively, while $T_2^{(\text{ub})}$ bounds the sum-rate in state 2 and in state 3.
\end{IEEEproof} 

\section{Quasi-Static Fading Channels}\label{sec:4}
\begin{figure*}[!b]
  \normalsize
  \setcounter{MYtempeqncnt}{\value{equation}}
  \setcounter{equation}{21}
  \vspace*{4pt}
  \hrulefill
  {\normalsize{
  \begin{IEEEeqnarray}{rl}\label{eq:RF_set1}
    \mathcal R_j\triangleq\Bigg\{
    &R_{1,j}\leq \log\det\rb{\mathbf{I}+\lambda_j P\mathbf{A}_1\sb{\lambda_2 P\rb{\mathbf{A}_1+\mathbf{A}_2}\mathds{1}_{\cb{j=1}}+\Text{diag}\rb{f_1(C_1),f_2(C_2)}}^{-1}},\IEEEnonumber\\
    &R_{2,j}\leq \log\det\rb{\mathbf{I}+\lambda_j P\mathbf{A}_2\sb{\lambda_2 P\rb{\mathbf{A}_1+\mathbf{A}_2}\mathds{1}_{\cb{j=1}}+\Text{diag}\rb{f_1(C_1),f_2(C_2)}}^{-1}},\IEEEnonumber\\
    &R_{1,j}+R_{2,j}\leq \log\det\rb{\mathbf{I}+\lambda_j P\rb{\mathbf{A}_1+\mathbf{A}_2}\sb{\lambda_2 P\rb{\mathbf{A}_1+\mathbf{A}_2}\mathds{1}_{\cb{j=1}}+\Text{diag}\rb{f_1(C_1),f_2(C_2)}}^{-1}}\Bigg\}
  \end{IEEEeqnarray}}}
  \setcounter{equation}{\value{MYtempeqncnt}}
\end{figure*}
In this section we study the scenario with quasi-static fading. While, as discussed in the previous section, the backhaul links have four possible states, the fading channels introduce an uncountable number of possible channel-backhaul states $\rb{\mathbf{a},\mathbf{c}}$. As a result, in principle, BC requires each MU to send an infinite number of layers to cope with the uncertainty on both fading channels and backhaul links (see \cite{Shamai}). However, works such as \cite{Liu} suggest that the full benefits of
BC can be often obtained with a very limited number of layers. Based on these results and aiming at reducing the complexity of the analysis, here we focus on two-layer BC ($K=2$). Therefore, each MU $j$ decomposes its message in two independent parts as $M_{j}=\rb{M_{j,1},M_{j,2}}$ with rates $R_{j,k}$, $k=1,2$, i.e., $M_{j,k}\in\cb{1,2,\ldots,2^{nR_{j,k}}}$. Each $j$th MU transmits (cf. \eqref{eq:NF_BC})
\begin{IEEEeqnarray}{rCl}\label{eq:RF_GB}
  X_{j}=\sqrt{P}\rb{\sqrt{\lambda_1}W_{j,1}+\sqrt{\lambda_2}W_{j,2}},
\end{IEEEeqnarray}
where $W_{j,k}\sim \mathcal C\mathcal N(0,1)$ for $k=1,2$ represent the codebooks corresponding to the two messages $M_{j,k}$, $k=1,2$, for MU $j=1,2$. Moreover, we consider compression at the BSs based on (complex) Gaussian test channels and successive refinement, similar to what was done for the non-fading case (cf. \eqref{eq:NF_test1} and \eqref{eq:NF_test2}). We recall that the BSs know the fading state and thus can adjust the compression noise variances to the current fading conditions. For simplicity of analysis, we consider only separate decompression, as in Proposition \ref{prop:achievable}, and leave the analysis of joint decompression and distributed source coding to future work.

Based on the compressed received signals of the BSs recovered at the RCP, the latter attempts decoding of the MUs' messages. Unlike the non-fading case, here the messages to be decoded are not determined by the backhaul state only (cf. Fig. \ref{fig:Drawing}), but also by the fading states. In order to assess which subset of messages $\rb{M_{1,1},M_{1,2},M_{2,1},M_{2,2}}$ are decodable in state $(\mathbf{a},\mathbf{c})$, we assume a successive decoding approach in which the RCP first attempts to decode jointly the first-layer messages $\rb{M_{1,1},M_{2,1}}$ and then, if the first-layer messages are both decoded correctly, it jointly decodes the second-layer messages $\rb{M_{1,2},M_{2,2}}$. We consider two specific approaches, whereby, in the first, messages at the same layer are assumed to be correctly decoded only if both messages at the same layer are; while, in the second, decoding of only one message per layer is allowed. We refer to the first approach as ``common outage'' decoding, while to the second as ``individual outage'' decoding. Note that ``common outage'' decoding imposes the stronger requirement that the messages of both MUs must be correctly received in order to declare the transmission of a layer successful. The propositions below provide achievable regions with these two approaches.
\begin{my_prop}[Common outage decoding]\label{prop:common}
Let $\lambda_1,\lambda_2\geq0$ such that $\lambda_1+\lambda_2=1$. The average throughput
\begin{IEEEeqnarray}{rCl}\label{eq:RF_throughput}
  T&=&\Pr\cb{\mathcal R_1}\rb{R_{1,1}+R_{2,1}}\IEEEnonumber\\
  && +\: \Pr\cb{\mathcal R_1\cap\mathcal R_2}\rb{R_{1,2}+R_{2,2}}
\end{IEEEeqnarray}
is achievable, where the sets $\mathcal R_1$ and $\mathcal R_2$ are defined as \eqref{eq:RF_set1} for $j=1,2$, at the bottom of the page, \addtocounter{equation}{1} with
\begin{IEEEeqnarray}{rCl}\label{eq:RF_def1}
  \mathbf{A}_1&\triangleq& \rb{\begin{array}{cc}\abs{a_{1,1}}^2&\alpha a_{1,1} a_{2,1}^*\\ \alpha a_{1,1}^* a_{2,1}&\alpha^2\abs{a_{2,1}}^2\end{array}},\IEEEyessubnumber\\
  \mathbf{A}_2&\triangleq& \rb{\begin{array}{cc}\alpha^2\abs{a_{1,2}}^2&\alpha a_{1,2} a_{2,2}^*\\ \alpha a_{1,2}^* a_{2,2}&\abs{a_{2,2}}^2\end{array}},\IEEEyessubnumber
\end{IEEEeqnarray}
and
\begin{IEEEeqnarray}{rCl}\label{eq:RF_def2}
  f_j(C_j)&=&\left\lbrace\begin{array}{ll}1+\sigma_{j,1}^2+\sigma_{j,2}^2&C_j=C\\1+\sigma_{j,2}^2&C_j=C+\Delta C\end{array}\right.
\end{IEEEeqnarray}
with
\begin{IEEEeqnarray}{rCl}\label{eq:RF_def3}\label{eq:RF_sigmas}
  \sigma_{j,1}^2&=&\frac{2^C\rb{2^{\Delta C}-1}}{\rb{2^C-1}}\cdot\sigma_{j,2}^2,\IEEEyessubnumber\\
  \sigma_{j,2}^2&=&\frac{P\rb{\abs{a_{j,j}}^2+\alpha^2\abs{a_{j,[j]_2+1}}^2}+1}{2^{C+\Delta C}-1}\IEEEyessubnumber.
\end{IEEEeqnarray}
\end{my_prop}
Proposition \ref{prop:common} is proved in Appendix \ref{app:RF_achievable} and is briefly described here.
The random throughput takes on two values, namely $R_{1,1}+R_{2,1}$ with probability $\Pr\cb{\mathcal R_1\cap\mathcal R_2^C}$, and $R_{1,1}+R_{1,2}+R_{2,1}+R_{2,2}$ with probability $\Pr\cb{\mathcal R_1\cap\mathcal R_2}$. Set $\mathcal R_1$ in \eqref{eq:RF_set1} for $j=1$ represents the subset of channel-backhaul states $(\mathbf{a},\mathbf{c})$ in which the first-layer messages of both MUs are jointly decodable. Similarly, the set $\mathcal R_2$ in \eqref{eq:RF_set1} for $j=2$ is the subset of channel-backhaul states $(\mathbf{a},\mathbf{c})$ for which the second-layer messages of the MUs are decodable when conditioning on the event that both first-layer messages have been decoded correctly. Thus, the average throughput is given by \eqref{eq:RF_throughput}. The regions $\mathcal R_1$ and $\mathcal R_2$ are obtained based on capacity results for multiple access channels \cite{KimGamal} as follows.

\begin{figure*}[!t]
  \normalsize
  \setcounter{MYtempeqncnt}{\value{equation}}
  \setcounter{equation}{26}
  {\normalsize{
  \begin{IEEEeqnarray}{rl}\label{RF_ind_R1}
    \mathcal R^{(j)}_1\triangleq\Bigg\{
    &R_{j,1}\leq \log\det\rb{\mathbf{I}+\lambda_1 P\mathbf{A}_j\sb{P\rb{\lambda_2\mathbf{A}_j+\mathbf{A}_{[j]_2+1}}+\Text{diag}\rb{f_1(C_1),f_2(C_2)}}^{-1}},\IEEEnonumber\\
    &R_{[j]_2+1,1}>\log\det\rb{\mathbf{I}+\lambda_1 P\mathbf{A}_{[j]_2+1}\sb{\lambda_2P\rb{\mathbf{A}_1+\mathbf{A}_2}+\Text{diag}\rb{f_1(C_1),f_2(C_2)}}^{-1}}\Bigg\}\IEEEeqnarraynumspace
  \end{IEEEeqnarray}}}
  {\normalsize{
  \begin{IEEEeqnarray}{rl}\label{RF_ind_R3}
    \mathcal R_{2+j}\triangleq\Bigg\{
    &R_{j,2}\leq \log\det\rb{\mathbf{I}+\lambda_2P\mathbf{A}_j \sb{\lambda_2P\mathbf{A}_{[j]_2+1}+\Text{diag}\rb{f_1(C_1),f_2(C_2)}}^{-1}},\IEEEnonumber\\
    &R_{[j]_2+1,2}> \log\det\rb{\mathbf{I}+\lambda_2P\mathbf{A}_{[j]_2+1} \sb{\Text{diag}\rb{f_1(C_1),f_2(C_2)}}^{-1}}\Bigg\}
  \end{IEEEeqnarray}}}
  {\normalsize{
  \begin{IEEEeqnarray}{rl}\label{RF_ind_R5}
    \mathcal R^{(j)}_2\triangleq\Bigg\{
    R_{j,2}\leq \log\det\rb{\mathbf{I}+\lambda_2P\mathbf{A}_j \sb{P\mathbf{A}_{[j]_2+1}+\Text{diag}\rb{f_1(C_1),f_2(C_2)}}^{-1}}\Bigg\}
  \end{IEEEeqnarray}}}
  \setcounter{equation}{\value{MYtempeqncnt}}
  \hrulefill
  \vspace*{4pt}
\end{figure*}
Similar to the strategy for the non-fading case, the MUs use i.i.d. generated complex Gaussian codebooks, as in \eqref{eq:RF_GB}, where parameters $\rb{\lambda_1,\lambda_2}$ represent the power allocation used by the $j$th MU to transmit the messages $\rb{M_{j,1},M_{j,2}}$. Moreover, at the RCP, we assume a successive decoding approach, whereby, in each decoding step, layers that are not decoded are considered additional noise. Separate compression at the BSs is performed, and $f_j(C_j)$ in \eqref{eq:RF_def2} represents the equivalent noise variance due to channel noise and compression noise that depends on the random state of the $j$th backhaul link and on the current fading gains. Specifically, as in the case without fading, the quantities $\rb{\sigma_{j,1}^2,\sigma_{j,2}^2}$ in \eqref{eq:RF_sigmas} represent the quantization noise introduced by compression at the BSs (cf. \eqref{eq:nf_sig}).

We now turn to the rate achievable with individual-outage decoding.
\begin{my_prop}[Individual outage decoding]\label{prop:individual}
Let $\lambda_1,\lambda_2\geq0$ such that $\lambda_1+\lambda_2=1$. The average throughput
{\small{
\begin{IEEEeqnarray}{rl}\label{eq:RF_ind_sum}
  T&=\\
  &\rb{\Pr\Big\{\mathcal R_1\Big\}+\Pr\cb{\mathcal R^{(1)}_1}}R_{1,1}+\rb{\Pr\Big\{\mathcal R_1\Big\}+\Pr\cb{\mathcal R^{(2)}_1}}R_{2,1}\IEEEnonumber\\
  &+ \:\rb{\Pr\Big\{\mathcal R_1\cap \mathcal R_2\Big\}+\Pr\Big\{\mathcal R_1\cap \mathcal R_3\Big\}+\Pr\cb{\mathcal R^{(1)}_1\cap \mathcal R^{(1)}_2}}R_{1,2}\IEEEnonumber\\
  &+ \:\rb{\Pr\Big\{\mathcal R_1\cap \mathcal R_2\Big\}+\Pr\Big\{\mathcal R_1\cap \mathcal R_4\Big\}+\Pr\cb{\mathcal R^{(2)}_1\cap \mathcal R^{(2)}_2}}R_{2,2}\IEEEnonumber
\end{IEEEeqnarray}}}
is achievable, with \eqref{eq:RF_set1}-\eqref{eq:RF_def3} and for $j=1,2$ $\mathcal R^{(j)}_1,\mathcal R_{2+j},\mathcal R^{(j)}_2$ defined as \eqref{RF_ind_R1}, \eqref{RF_ind_R3}, \eqref{RF_ind_R5}, respectively, at the top of the next page. \addtocounter{equation}{3}
\end{my_prop}
As mentioned above, with individual outage decoding, we allow the RCP to decode only one of the messages of the two MUs at each layer. As a result, the throughput takes on eight possible values:
\begin{itemize}
\item $R_{j,1}+R_{j,2}$ with probability $\Pr\cb{\mathcal R^{(j)}_1\cap\mathcal R^{(j)}_2}$, for $j=1,2$. This throughput value corresponds to decoding first and second layer messages only of MU $j=1,2$. The set $\mathcal R^{(j)}_1$ in \eqref{RF_ind_R1} represents the conditions needed to correctly decode the first-layer message $M_{j,1}$ considering all the other messages as additional noise, and $\mathcal R^{(j)}_2$ in \eqref{RF_ind_R5} represents the conditions needed to correctly decode the second-layer message $M_{j,2}$ conditioned on message $M_{j,1}$ having already been decoded, considering the remaining undecoded messages as noise.
\item $R_{j,1}$ with probability $\Pr\cb{\mathcal R^{(j)}_1\cap\rb{\mathcal R^{(j)}_2}^C}$. This value corresponds to decoding only the first-layer message of MU $j=1,2$.
\item $R_{1,1}+R_{2,1}+R_{1,2}+R_{2,2}$ with probability $\Pr\cb{\mathcal R_1\cap\mathcal R_2}$. This value corresponds to decoding all the messages of all MUs.
\item $R_{1,1}+R_{2,1}+R_{j,2}$ with probability $\Pr\cb{\mathcal R_1\cap \mathcal R_{2+j}}$. This value corresponds to decoding the first-layer messages of both MUs and the second-layer message only of MU $j=1,2$. The set $\mathcal R_{2+j}$ in \eqref{RF_ind_R3} represents the conditions under which the second-layer message $M_{j,2}$ is decodable once both first-layer messages $\rb{M_{1,1},M_{2,1}}$ have been already decoded.
\item $R_{1,1}+R_{2,1}$ with probability $\Pr\cb{\mathcal R_1\cap\mathcal R_2^C\cap\mathcal R_3^C\cap\mathcal R_4^C}$. This value corresponds to decoding only the first-layer messages of both MUs, no second-layer message gets decoded.
\end{itemize}
Thus, the average throughput is given by \eqref{eq:RF_ind_sum}. Further details can be found in Appendix \ref{app:RF_achievable}. 

\section{Numerical Results}\label{sec:5}
Here we present numerical results in order to gain insight into the performance of the robust strategies presented in the previous sections for both the non-fading and the fading scenarios.
\subsection{No Fading}
We start by assessing the impact of BC and layered compression on the performance with no fading using Proposition \ref{prop:achievable} and Proposition \ref{cor:achievable}. To this end, we compare the performance of the proposed scheme with five layers with special cases consisting of a reduced number of layers per user. Specifically, we consider a one layer strategy with $\lambda_1=1$, two two-layer strategies, namely scheme 1 with $\lambda_1+\lambda_5=1$ and scheme 2 with $\lambda_1+\lambda_2=1$, and a three layer strategy with $\lambda_1+\lambda_2+\lambda_5=1$. In most considered instances, we found no significant gain in using the four-layer strategy $\rb{\lambda_1+\lambda_2+\lambda_4+\lambda_5=1}$ over the three layer strategy, and hence this strategy is not presented. We also compare the performance of separate and joint decompression corresponding to the performance characterized in Proposition \ref{prop:achievable} and Proposition \ref{cor:achievable}, respectively, along with the upper bound of Proposition \ref{prop:upper}. Note that in the following figures the average throughput is optimized over the power allocation parameters.
\begin{figure}[!t]
\centering
\includegraphics[scale=0.4]{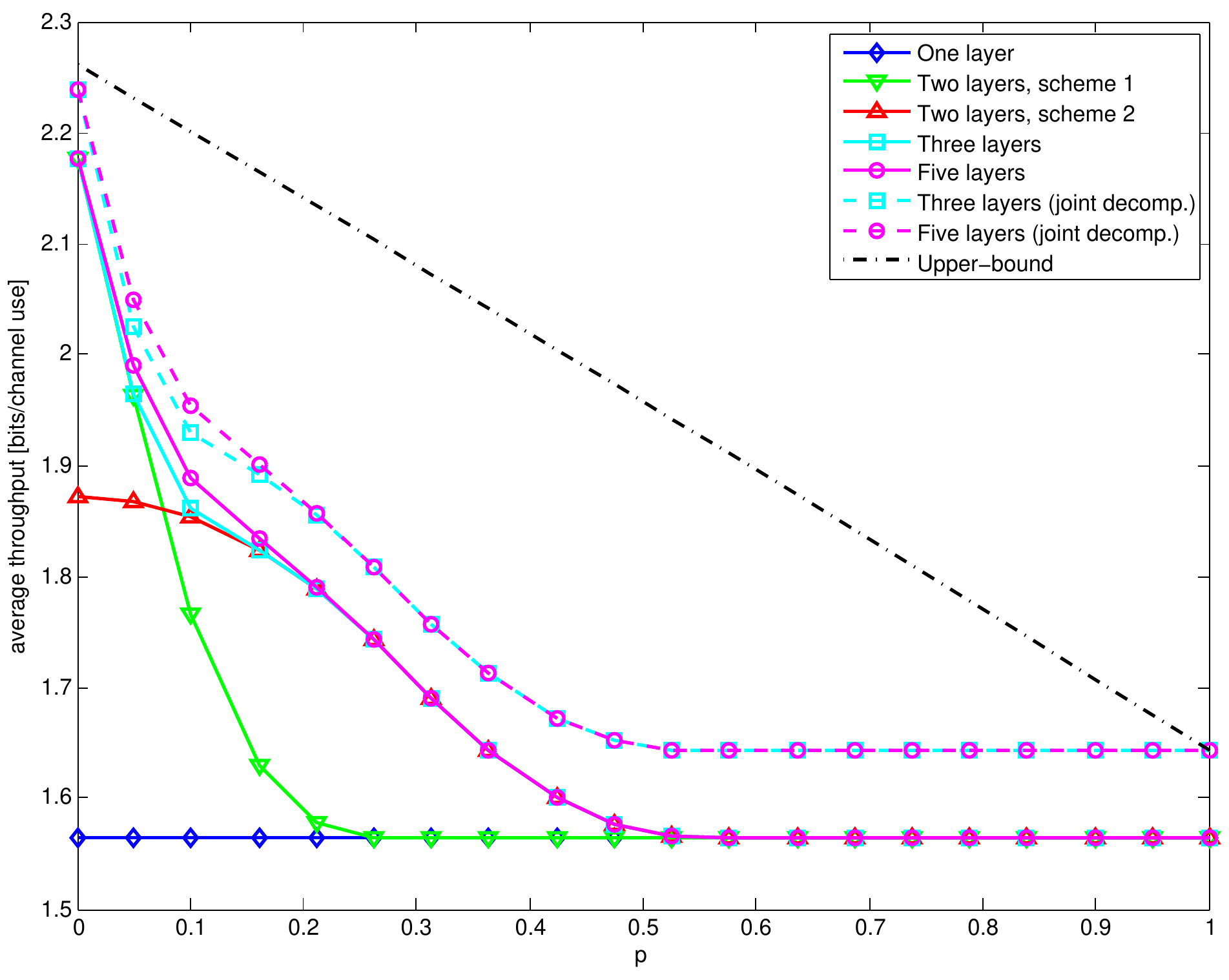}
\caption{Average throughput $T$ versus $p$ for $P=10dB,\alpha=0.3, C=1$ bits/channel use and $\Delta C=0.5$ bits/channel use.}
\label{fig:Rvsp}
\end{figure}

Fig. \ref{fig:Rvsp} shows the average throughput versus the probability $p$ of the each backhaul link to be in the low-capacity state $C_j=C$ for $P=10$dB, $\alpha=0.3$, $C=1$ bits/channel use and $\Delta C=0.5$ bits/channel use.
We first observe that increasing the number of layers leads to relevant throughput gains and that the same is true of joint decompression versus separate compression. In particular, the gain of joint decompression can be seen by comparing the achievable throughput performance for sufficiently large $p$. We also note that, for $p$ large enough, using a larger number of layers is not advantageous. This is because, when $p$ is large, the backhaul tends to be in the low-capacity state and higher layers mostly cause a decrease in signal-to-noise ratio in the states where they are not decoded.
\begin{figure}[!t]
\centering
\includegraphics[scale=0.4]{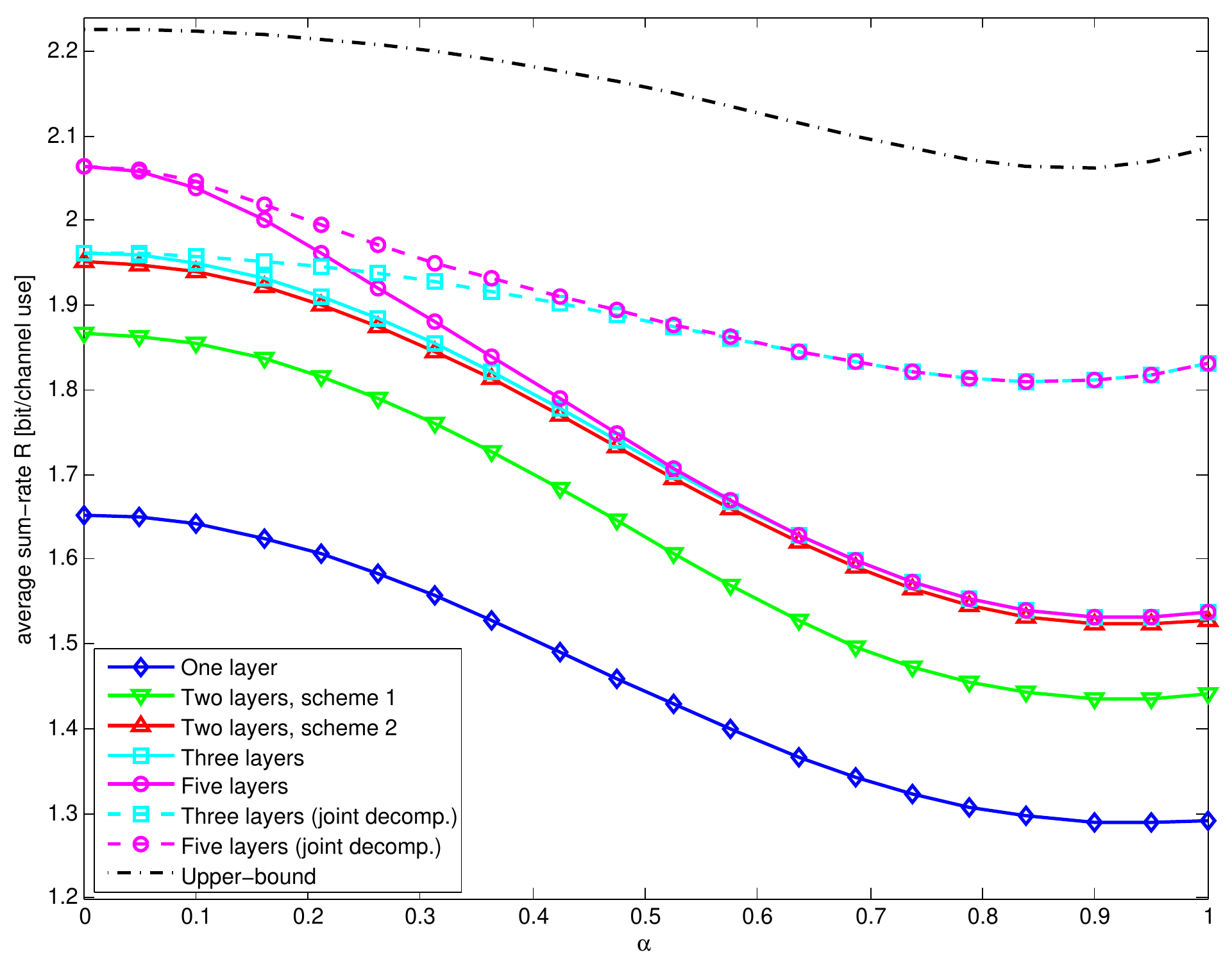}
\caption{Average throughput $T$ versus the inter-cell interference, $\alpha$ for $P=10dB,p=0.1,C=1$ bits/channel use and $\Delta C=0.5$ bits/channel use.}
\label{fig:RvsAlpha1}
\end{figure}

Fig. \ref{fig:RvsAlpha1} shows the average throughput versus the inter-cell interference factor $\alpha$ for $P=10$dB, $C=1$ bits/channel use, $\Delta C=0.5$ bits/channel use and $p=0.1$.
The performance gains observed above for schemes that use a larger number of layers and joint decompression is confirmed. In particular, it is interesting to note that the performance gain of joint decompression become more pronounced as $\alpha$ increases due to the larger correlation between the signals received by the BSs.
\begin{figure}[!t]
\centering
\includegraphics[scale=0.4]{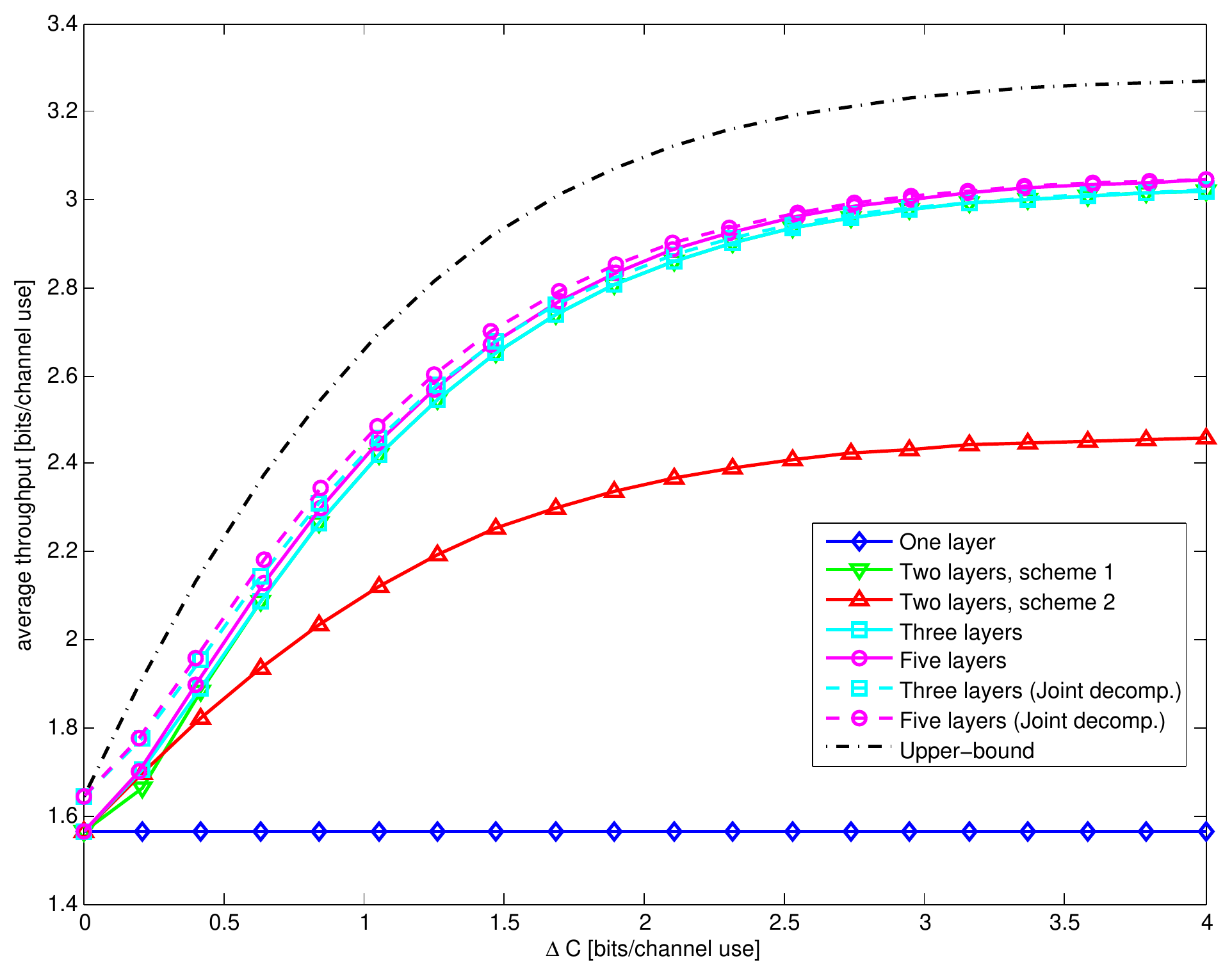}
\caption{Average throughput $T$ versus $\Delta C$ for $P=10dB,\alpha=0.3,p=0.05$ and $C=1$ bits/channel use.}
\label{fig:RvsDeltaC}
\end{figure}

Fig. \ref{fig:RvsDeltaC} shows the average throughput versus the capacity $\Delta C$ for $P=10$dB, $C=1$ bits/channel use, $\alpha=0.3$ and $p=0.05$.
As $\Delta C$ increases, there is a growing gap between the second two-layer strategy and the other multi-layer strategies. The reason for this behavior is that for small $p$, states where only one backhaul link is of high-capacity are rare compared to the state were both backhaul links are of high-capacity. Hence, strategies that allocate power to the fifth layer ($\lambda_5>0$), which gets decoded when $C_1=C_2=C+\Delta C$, result in a larger throughput compared to the second two-layer strategy where $\lambda_5=0$.

\subsection{Quasi-Static Fading}
We now turn to the performance achievable in the presence of quasi-static fading. The main goal is that of comparing the performance achievable with BC as a function of the number of layers and also of assessing the impact of the common and individual outage approaches. Note that the common outage performance is obtained from Proposition \ref{prop:common}, while that of the more complex individual outage strategy from Proposition \ref{prop:individual}. For both decoding schemes, one-layer ($\lambda_1=1$) and two-layer ($\lambda_1,\lambda_2>0$) strategies are considered. Note that in the following figures the average throughput is optimized over the power allocation parameters and choice of rates $\cb{R_{j,1},R_{j,2}}_{j=1}^{2}$. In addition, the upper bound obtained with full channel and backhaul state information is not shown given that it is not tight enough.
\begin{figure}[!t]
\centering
\includegraphics[scale=0.4]{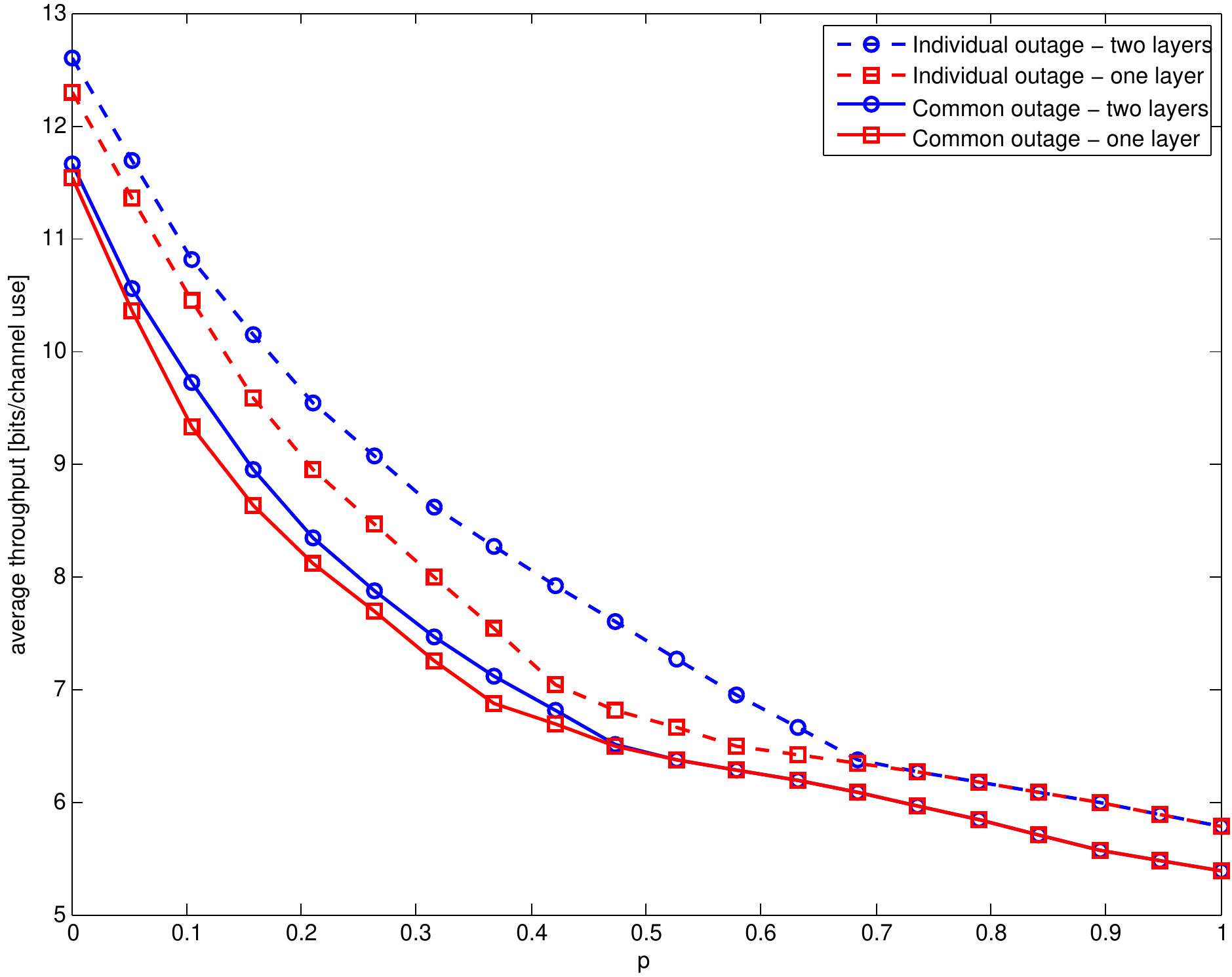}
\caption{Average throughput $T$ versus $p$ for $P=30dB,\alpha=0.3,C=4$ bits/channel use and $\Delta C=6$ bits/channel use.}
\label{fig:fading_Rvsp1}
\end{figure}

Figure \ref{fig:fading_Rvsp1} shows the average throughput versus the probability $p$ of each backhaul link to have capacity $C$ for $P=30dB,\alpha=0.3, C=4$ bits/channel use and $\Delta C=6$ bits/channel use. As expected,
individual-outage based decoding outperforms common-outage based decoding and one-layer strategies are outperformed by two-layer strategies. Note that the performance gain of  BC, i.e., of using two layers, is apparent even when $p=0$, that is, when no backhaul link uncertainty occurs. This is because BC still allows the negative effects of the uncertainty about the fading channels to be alleviated.
It is also noted that, similar to the non-fading case of Fig. \ref{fig:Rvsp}, for $p$ large enough, no performance gain is accrued by using BC. In fact, while BC can still be useful in combating fading, when $p$ is large, the backhaul is often in the low-capacity state. Therefore, in this situation the noise due to compression dominates the performance and BC leads to negligible gains. These results are consistent with the known small advantage of BC in the low signal-to-noise ratio regime  (\cite{Shamai},\cite{Liu}).
\begin{figure}[!t]
\centering
\includegraphics[scale=0.4]{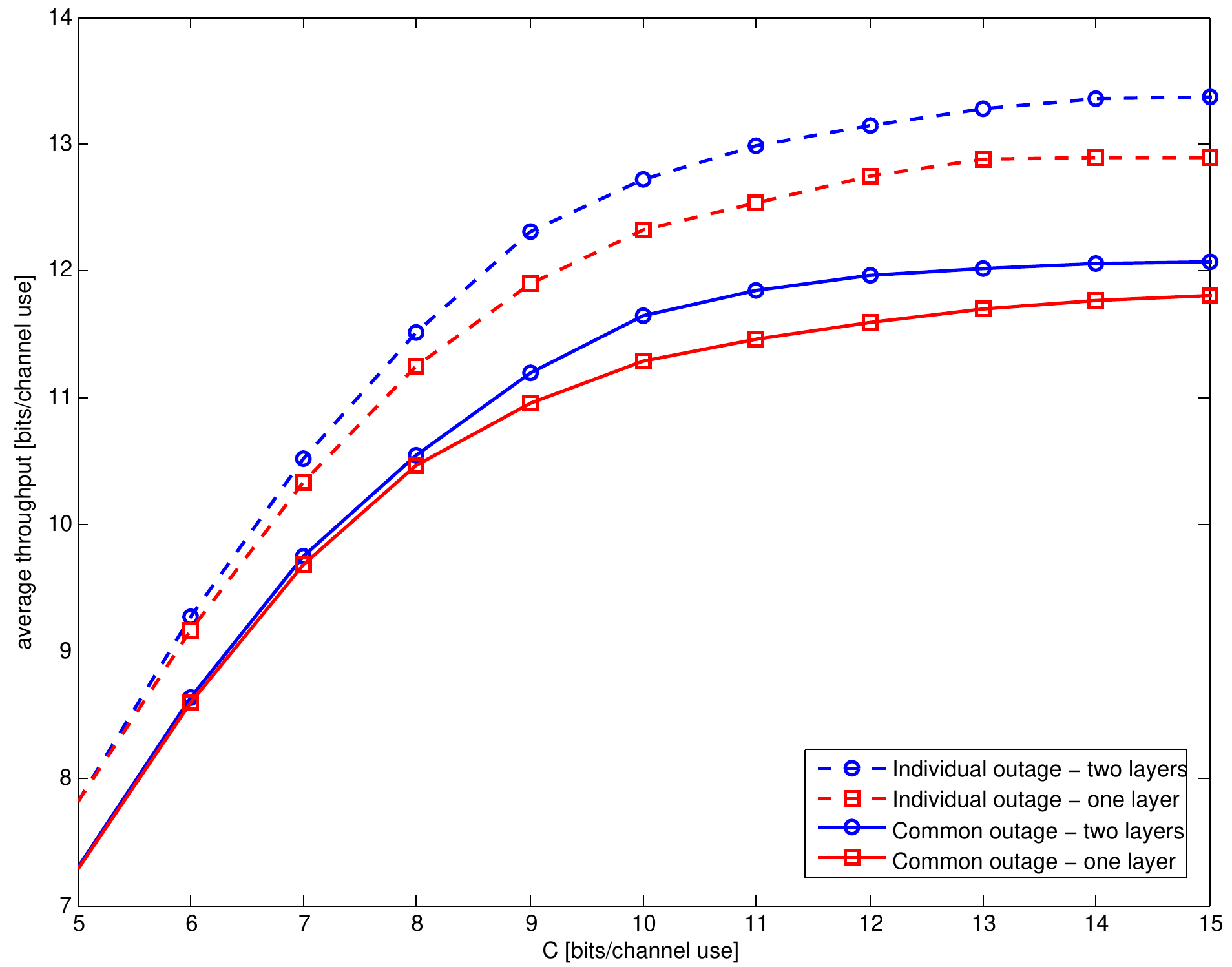}
\caption{Average throughput $T$ versus the backhaul link capacity $C$ in the low-capacity state for $P=30dB,\alpha=0.2$, and $\Delta C=0$ bits/channel use.}
\label{fig:fading_RvsC1}
\end{figure}

Following on the discussion above, we now observe how the performance gains of BC depend on the backhaul capacity $C$. To this end, Figure \ref{fig:fading_RvsC1} shows the average throughput versus the backhaul capacity $C$ for $\alpha=0.2$, $P=30dB$, and $\Delta C=0$ bits/channel use. As discussed above, for small $C$, there is no gain in using more than one layer since the compression noise dominates the performance. However, as $C$ increases and thus the effect of the compression noise decreases, BC outperforms the single-layer strategy due to its robust operation with respect to the uncertainty over the fading channels.
\begin{figure}[!t]
\centering
\includegraphics[scale=0.4]{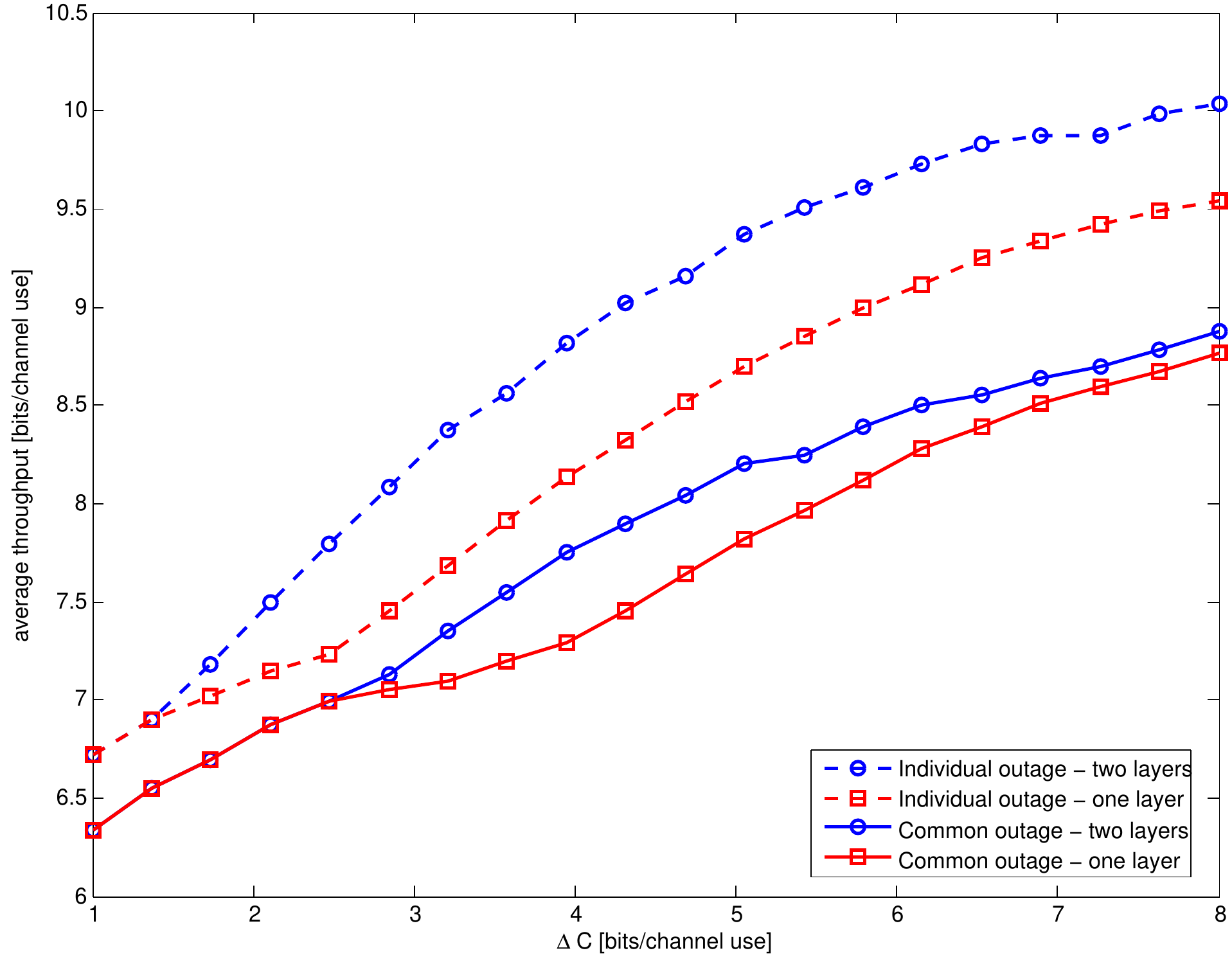}
\caption{Average throughput $T$ versus $\Delta C$ for $P=30dB, \alpha=0.3 ,p=0.2$ and $C=4$ bits/channel use.}
\label{fig:fading_RvsDeltaC}
\end{figure}

Finally, Figure \ref{fig:fading_RvsDeltaC} shows the average throughput versus the capacity $\Delta C$ for $P=30dB$, $\alpha=0.3$, $C=4$ bits/channel use and $p=0.2$.
It is interesting to remark that the performance gain of the BC scheme with respect to transmission of one-layer decreases for increasing $\Delta C$. This is because, when $\Delta C$ is much larger than $C$ and $p$ is small enough, the gain of the BC strategy only arises from the possibility of combating fading and not from dealing with the uncertainty of the backhaul links.

\section{Concluding Remarks}\label{sec:6}
In delay-constrained applications, it is often unrealistic to assume transmit channel state information. In a cloud radio access system, this calls for robust transmission strategies both in the first hop between the MUs and the BSs and in the second hop consisting of the backhaul links between BSs and RCP. In this paper, we have proposed such a robust transmission strategy based on BC at the MUs and layered compression at the BSs. The analysis and numerical results reveal the importance of BC and layered compression, especially when coupled with distributed source coding, in opportunistically leveraging advantageous channel and backhaul conditions. Future interesting work includes the analysis of systems with more than two cells and the investigation of joint decompression and decoding strategies as in \cite{Sanderovich2}.

\appendices
\section{Proof of Propositions \ref{prop:achievable} and \ref{cor:achievable}}\label{app:NF_achievable}
\begin{figure*}[!t]
  \normalsize
  \setcounter{MYtempeqncnt}{\value{equation}}
  \setcounter{equation}{32}
  {\normalsize{
  \begin{IEEEeqnarray}{rCl}\label{eq:appA_MAC}
    R_{1,1}+R_{2,1}&\leq&I(W_{1,1},W_{2,1};V_{1,1},V_{2,1}),\IEEEyessubnumber\\
    R_{1,2}+R_{2,2}&\leq&I(W_{1,2},W_{2,2};V_{j,2},V_{[j]_2+1,1}|W_{1,1},W_{2,1}),\IEEEyessubnumber\\
    R_{j,3}&\leq&I(W_{j,3};V_{j,2},V_{[j]_2+1,1}|W_{j,2},W_{[j]_2+1,4}),\IEEEyessubnumber\\
    R_{j,4}&\leq&I(W_{j,4};V_{[j]_2+1,2},V_{j,1}|W_{[j]_2+1,3},W_{j,2}),\IEEEyessubnumber\\
    R_{j,3}+R_{[j]_2+1,4}&\leq&I(W_{j,3},W_{[j]_2+1,4};V_{j,2},V_{[j]_2+1,1}|W_{j,2},W_{[j]_2+1,2}),\IEEEyessubnumber\\
    R_{1,5}+R_{2,5}&\leq&I(X_1,X_2;V_{1,2},V_{2,2}|W_{1,3},W_{1,4},W_{2,3},W_{2,4}).\IEEEyessubnumber
  \end{IEEEeqnarray}
  }}
  \setcounter{equation}{\value{MYtempeqncnt}}
  \hrulefill
  \vspace*{4pt}
\end{figure*}
Each MU uses random i.i.d. Gaussian codebooks generated according to \eqref{eq:NF_BC}. Therefore, each decoder in Fig. \ref{fig:Drawing}. sees a multiple access channel with a different number of transmitters and effective noise levels. For instance, Decoder 1 sees a multiple access channel with two users with inputs $W_{1,1}$ and $W_{2,1}$, while the effective noise levels include all other transmitted layers $W_{j,k}$ with $k\neq1$. The conditions on the rates $R_{j,k}$, for $j=1,2$ and $k=1,2,\ldots,5$, then follow from standard results on the capacity region of multiple access channels (see, e.g., \cite[Ch. 4.1]{KimGamal}), once one identifies the signals received at each decoder.

To this end, recall that, with separate decompression, each BS performs successive refinement quantization with test channels \eqref{eq:NF_test1}-\eqref{eq:NF_test2}. From standard results \cite[Ch. 13.5]{KimGamal}, we have that the conditions
\begin{IEEEeqnarray}{rCl}\label{eq:appA_C}
  C&\geq&I\rb{Y_j;V_{j,1}}
\end{IEEEeqnarray}
and
\begin{IEEEeqnarray}{rCl}\label{eq:appA_DC}
    \Delta C&\geq&I\rb{\given{Y_j;V_{j,2}}{V_{j,1}}}
\end{IEEEeqnarray}
are sufficient for the $j$th BS to convey the coarse compressed version $V_{j,1}$ in \eqref{eq:NF_test1} to the RCP over the backhaul link of capacity $C$ and the refined description $V_{j,2}$ in \eqref{eq:NF_test2} when the $j$th backhaul link is in state $C+\Delta C$. Imposing equality in \eqref{eq:appA_C}-\eqref{eq:appA_DC} leads to \eqref{eq:nf_sig}.

Alternatively, with joint decompression of the coarse descriptions, the conditions \eqref{eq:appA_C} for $j=1,2$ can be alleviated to the sum-rate constraint
\begin{IEEEeqnarray}{rCl}\label{eq:appA_2C}
  2C&\geq& H\rb{V_{1,1},V_{2,1}}-\:H\rb{V_{1,1}|Y_1}-H\rb{V_{2,1}|Y_2},
\end{IEEEeqnarray}
leading to \eqref{eq:nf_sig3}. We observe that this result hinges on the symmetry of the system model as in \cite{Simeone}.

We can now identify the received signals by each decoder in Fig. \ref{fig:Drawing}. For instance, Decoder 1 receives $V_{1,1}$ and $V_{2,1}$. Using capacity results on multiple access channels \cite[Ch. 4.1]{KimGamal} and the symmetry of the system, we get the conditions \addtocounter{equation}{1} \eqref{eq:appA_MAC} at the top of the next page for correct decoding at all decoders.
Evaluating these conditions with \eqref{eq:NF_BC}-\eqref{eq:NF_test2} completes the proof.

\section{Proof of Propositions \ref{prop:common} and \ref{prop:individual}}\label{app:RF_achievable}
As seen in Sec. \ref{sec:4}, the MUs use two-layer BC with random i.i.d. complex Gaussian codebooks generated according to \eqref{eq:RF_GB}. The $j$th BS receives $Y_j^n$ from the fading channel, and performs successive refinement quantization with test channels \eqref{eq:NF_test1} and \eqref{eq:NF_test2}, where $Q_{j,1}\sim \mathcal C\mathcal N(0,\sigma_{j,1}^2)$ and $Q_{j,2}\sim \mathcal C\mathcal N(0,\sigma_{j,2}^2)$. We recall that the BSs know the fading state and thus can adjust the compression noise variances to the current fading conditions. As was mentioned, the focus here is only on separate decompression, and hence, similar to the non-fading case, \eqref{eq:appA_C} and \eqref{eq:appA_DC} are sufficient for the $j$th BS to convey the coarse compressed version $V_{j,1}$ or the refined description $V_{j,2}$ to the RCP over backhaul link of capacity $C$ or $C+\Delta C$, respectively. Imposing equality in \eqref{eq:appA_C}-\eqref{eq:appA_DC} leads to \eqref{eq:RF_sigmas}.

Denote by $\mathcal V_{c_1,c_2}$ the received signals by the RCP in backhaul links' state $(c_1,c_2)$, i.e., $\mathcal V_{C,C}=\rb{V_{1,1},V_{2,1}}$, $\mathcal V_{C+\Delta C,C}=\rb{V_{1,2},V_{2,1}}$, $\mathcal V_{C,C+\Delta C}=\rb{V_{1,1},V_{2,2}}$ and $\mathcal V_{C+\Delta C,C+\Delta C}=\rb{V_{1,2},V_{2,2}}$. The RCP is aware of the channel-backhaul state, and uses a successive decoding approach. Based on standard results on the capacity region of multiple access channels (see, e.g., \cite[Ch. 4.1]{KimGamal}), the RCP can jointly decode both first-layer messages, in channel-backhaul state $\rb{\mathbf{a},\mathbf{c}}$, if $\rb{R_{1,1},R_{2,1}}\in\mathcal R_1$ where
\begin{IEEEeqnarray}{rl}\label{eq:appC_set1}
  \mathcal R_1=\Bigg\{
  &R_{1,1}\leq I\rb{\given{W_{1,1};\mathcal V_{c_1,c_2}}{W_{2,1}}},\IEEEnonumber\\
  &R_{2,1}\leq I\rb{\given{W_{2,1};\mathcal V_{c_1,c_2}}{W_{1,1}}},\IEEEnonumber\\
  &R_{1,1}+R_{2,1}\leq I\rb{W_{1,1},W_{2,1};\mathcal V_{c_1,c_2}} \Bigg\},\IEEEeqnarraynumspace
\end{IEEEeqnarray}
which together with \eqref{eq:NF_test1}, \eqref{eq:NF_test2} and \eqref{eq:RF_GB} leads to \eqref{eq:RF_set1} for $j=1$. In common outage decoding, if $\rb{R_{1,1},R_{2,1}}\notin\mathcal R_1$ an outage is declared. However, in individual outage decoding we allow scenarios where only one message can be decoded. Notice that the RCP can decode the first-layer message of the $j$th MU (but not the first-layer message of the other user), in channel-backhaul state $\rb{\mathbf{a},\mathbf{c}}$, if $\rb{R_{j,1},R_{[j]_2+1,1}}\in\mathcal R^{(j)}_1$ where
\begin{IEEEeqnarray}{rl}\label{eq:appC_set3}
  \mathcal R^{(j)}_1=\Bigg\{
  &R_{j,1}\leq I\rb{W_{j,1};\mathcal V_{c_1,c_2}},\IEEEnonumber\\
  &R_{[j]_2+1,1}>I\rb{\given{W_{[j]_2+1,1};\mathcal V_{c_1,c_2}}{W_{j,1}}}\Bigg\},\IEEEeqnarraynumspace
\end{IEEEeqnarray}
which together with \eqref{eq:NF_test1}, \eqref{eq:NF_test2} and \eqref{eq:RF_GB} leads to \eqref{RF_ind_R1}. As for the second-layer messages, assuming both first-layer messages were decoded, the RCP can jointly decode both second-layer messages, in channel-backhaul state $\rb{\mathbf{a},\mathbf{c}}$, if $\rb{R_{2,1},R_{2,2}}\in\mathcal R_2$ where
\begin{IEEEeqnarray}{rl}\label{eq:appC_set2}
  \mathcal R_2=&\IEEEnonumber\\\Bigg\{
  &R_{1,2}\leq I\rb{\given{X_1;\mathcal V_{c_1,c_2}}{W_{1,1},X_2}},\IEEEnonumber\\
  &R_{2,2}\leq I\rb{\given{X_2;\mathcal V_{c_1,c_2}}{W_{2,1},X_1}},\IEEEnonumber\\
  &R_{1,2}+R_{2,2}\leq I\rb{\given{X_1,X_2;\mathcal V_{c_1,c_2}}{W_{1,1},W_{2,1}}} \Bigg\},\IEEEeqnarraynumspace
\end{IEEEeqnarray}
which together with \eqref{eq:NF_test1}, \eqref{eq:NF_test2} and \eqref{eq:RF_GB} leads to \eqref{eq:RF_set1} for $j=2$. Otherwise, with common outage decoding a second-layer outage is declared, while with individual outage decoding, similarly to the first layer, the RCP can decode the second-layer message of the $j$th MU (but not the second-layer message of the other user), in channel-backhaul state $\rb{\mathbf{a},\mathbf{c}}$, if $\rb{R_{j,2},R_{[j]_2+1,2}}\in\mathcal R_{j+2}$ where
\begin{IEEEeqnarray}{rl}\label{eq:appC_set4}
  \mathcal R_{2+j}=&\IEEEnonumber\\\Bigg\{
  &R_{j,2}\leq I\rb{\given{X_j;\mathcal V_{c_1,c_2}}{W_{1,1},W_{2,1}}},\IEEEnonumber\\
  &R_{[j]_2+1,2}> I\rb{\given{X_{[j]_2+1};\mathcal V_{c_1,c_2}}{W_{[j]_2+1,1},X_j}}\Bigg\},\IEEEeqnarraynumspace
\end{IEEEeqnarray}
which together with \eqref{eq:NF_test1}, \eqref{eq:NF_test2} and \eqref{eq:RF_GB} leads to \eqref{RF_ind_R3}. Instead, if only the first-layer message of the $j$th MU was decoded, then the RCP can decode also the second-layer message of the $j$th MU, in channel-backhaul state $\rb{\mathbf{a},\mathbf{c}}$, if $R_{j,2}\in\mathcal R^{(j)}_2$ where
\begin{IEEEeqnarray}{rl}\label{eq:appC_set5}
  \mathcal R^{(j)}_2=\Bigg\{
  R_{j,2}\leq I\rb{\given{X_j;\mathcal V_{c_1,c_2}}{W_{j,1}}}\Bigg\},
\end{IEEEeqnarray}
which together with \eqref{eq:NF_test1}, \eqref{eq:NF_test2} and \eqref{eq:RF_GB} leads to \eqref{RF_ind_R5}. Thus, the average throughput for common outage decoding is given by \eqref{eq:RF_throughput}, while for individual outage decoding, the average throughput is given by \eqref{eq:RF_ind_sum}.

\ifCLASSOPTIONcaptionsoff
  \newpage
\fi

\bibliographystyle{IEEEtran}
\bibliography{IEEEabrv,main}

\end{document}